\documentclass[twocolumn]{aastex61}
\usepackage{amsmath}
\usepackage{booktabs}
\usepackage[utf8]{inputenc}
\usepackage{xcolor}
\usepackage{graphicx}

\newcommand{\CoincidentTime}{48.6 days} 
\newcommand{\TotalSingleTime}{44.5 days} 
\newcommand{\TotalLiveTime}{93.2 days} 

\newcommand{\SearchRange}{$\sim85 \, \text{Mpc}$}
\newcommand{\NumSingleCandidates}{15}
\newcommand{\NumCandidates}{103}
\newcommand{\OverlapOGC}{15}

\newcommand{\VT}{$6.7 \times 10^{5} \, \mathrm{Mpc}^3 \, \mathrm{yr}$} 
\newcommand{\ProbedVT}{$0.77 \, \mathrm{Gpc}^{3} \, \mathrm{yr}$} 

\newcommand{\NumInjections}{$112\thinspace073$}
\newcommand{\NumRejections}{$17\thinspace738\thinspace506$}

\newcommand{\BNSRate}{\ensuremath{\mathcal{R} \approx 100 \, - 4000 \, \mathrm{Gpc}^{-3} \, \mathrm{yr}^{-1}}}

\newcommand{\WebsiteURL}{https://dcc.ligo.org/public/0158/P1900030/001/index.html}

\newcommand{\ObsOneStart}{September 12, 2015}
\newcommand{\ObsOneEnd}{January 19, 2016}

\newcommand{\BankSize}{65\thinspace634}
\newcommand{\BankMM}{99\%}

\newcommand{\likelihood}{\ensuremath{\mathcal{L}}}

\begin{document}

\title{Sub-threshold binary neutron star search in advanced LIGO's first observing run}

\author{Ryan Magee}
\affiliation{Department of Physics, The Pennsylvania State University, University Park, PA 16802, USA}
\affiliation{Institute for Gravitation and the Cosmos, The Pennsylvania State University, University Park, PA 16802, USA}

\author{Heather Fong}
\affiliation{Canadian Institute for Theoretical Astrophysics, 60 St. George Street, University of Toronto, Toronto, Ontario, M5S 3H8, Canada}
\affiliation{Department of Physics, 60 St. George Street, University of Toronto, Toronto, Ontario, M5S 3H8, Canada}
\affiliation{RESCEU, The University of Tokyo, Tokyo, 113-0033, Japan}

\author{Sarah Caudill}
\affiliation{Nikhef, Science Park, 1098 XG Amsterdam, Netherlands}

\author{Cody Messick}
\affiliation{Department of Physics, The Pennsylvania State University, University Park, PA 16802, USA}
\affiliation{Institute for Gravitation and the Cosmos, The Pennsylvania State University, University Park, PA 16802, USA}

\author{Kipp Cannon}
\affiliation{Canadian Institute for Theoretical Astrophysics, 60 St. George Street, University of Toronto, Toronto, Ontario, M5S 3H8, Canada}
\affiliation{RESCEU, The University of Tokyo, Tokyo, 113-0033, Japan}

\author{Patrick Godwin}
\affiliation{Department of Physics, The Pennsylvania State University, University Park, PA 16802, USA}
\affiliation{Institute for Gravitation and the Cosmos, The Pennsylvania State University, University Park, PA 16802, USA}

\author{Chad Hanna}
\affiliation{Department of Physics, The Pennsylvania State University, University Park, PA 16802, USA}
\affiliation{Department of Astronomy and Astrophysics, The Pennsylvania State University, University Park, PA 16802, USA}
\affiliation{Institute for Gravitation and the Cosmos, The Pennsylvania State University, University Park, PA 16802, USA}
\affiliation{Institute for CyberScience, The Pennsylvania State University, University Park, PA 16802, USA}

\author{Shasvath Kapadia}
\affiliation{Leonard E.\ Parker Center for Gravitation, Cosmology, and Astrophysics, University of Wisconsin-Milwaukee, Milwaukee, WI 53201, USA}

\author{Duncan Meacher}
\affiliation{Leonard E.\ Parker Center for Gravitation, Cosmology, and Astrophysics, University of Wisconsin-Milwaukee, Milwaukee, WI 53201, USA}

\author{Siddharth R. Mohite}
\affiliation{Leonard E.\ Parker Center for Gravitation, Cosmology, and Astrophysics, University of Wisconsin-Milwaukee, Milwaukee, WI 53201, USA}
\affiliation{LSSTC Data Science Fellow}

\author{Debnandini Mukherjee}
\affiliation{Leonard E.\ Parker Center for Gravitation, Cosmology, and Astrophysics, University of Wisconsin-Milwaukee, Milwaukee, WI 53201, USA}

\author{Alexander Pace}
\affiliation{Department of Physics, The Pennsylvania State University, University Park, PA 16802, USA}

\author{Surabhi Sachdev}
\affiliation{Department of Physics, The Pennsylvania State University, University Park, PA 16802, USA}
\affiliation{Institute for Gravitation and the Cosmos, The Pennsylvania State University, University Park, PA 16802, USA}

\author{Minori Shikauchi}
\affiliation{Department of Astronomy, School of Science, the University of Tokyo, Hongo, Tokyo 113-0033, Japan}

\author{Leo Singer}
\affiliation{NASA/Goddard Space Flight Center, Greenbelt, MD 20771, USA}

\begin{abstract} We present a search for gravitational waves from double neutron star binaries
inspirals in Advanced LIGO's first observing run.  The search considers a
narrow range of binary chirp masses motivated by the population of known double
neutron star binaries in the nearby universe.  This search differs from
previously published results by providing the most sensitive published survey
of neutron stars in Advanced LIGO's first observing run within this narrow mass
range and including times when only one of the two LIGO detectors was in
operation in the analysis. The search was sensitive to binary neutron star
inspirals to an average distance of \SearchRange{} over \TotalLiveTime{}.  We
do not identify any unambiguous gravitational wave signals in our sample of
\NumCandidates{} sub-threshold candidates with false-alarm-rates of less than
one per day.  However, given the expected binary neutron star merger rate of
\BNSRate{}, we expect $\mathcal{O}(1)$ gravitational wave events within our
candidate list. This suggests the possibility that one or more of these
candidates is in fact a binary neutron star merger.  Although the contamination
fraction in our candidate list is $\sim99\%$, it might be possible to
correlate these events with other messengers to identify a potential
multi-messenger signal. We provide an online candidate list with the times and
sky locations for all events in order to enable multi-messenger searches.
\end{abstract} 


\section{Introduction}

Advanced LIGO~\citep{TheLIGOScientific:2014jea} conducted its first observing run (O1) from \ObsOneStart{} to
\ObsOneEnd{}.  Previous analyses of the 51.5 days of coincident LIGO Hanford
and LIGO Livingston data led to three detections of binary black hole
mergers~\citep{Abbott:2016blz, Abbott:2016nmj, TheLIGOScientific:2016pea,
LIGOScientific:2018mvr, nitz20181}.  No binary neutron star (BNS) or
neutron-star, black-hole (NSBH) systems were observed~\citep{Abbott:2016ymx} in
O1. We revisit this data with a gravitational wave search targeted at binary neutron star masses and provide a list of
candidate events. Searches that catalog low signal-to-noise ratio
(SNR) events probe significantly deeper into the cosmos. At low SNR it can be
difficult to claim an unambiguous detection, but the multi-messenger nature of
BNS systems~\citep{GBM:2017lvd} can be leveraged to identify authentic
gravitational wave events. Comparisons of catalogs provide a
discovery space for a host of multi-messenger
signals~\citep{smith2013astrophysical,Burns:2018pcl}. Temporal and/or spatial
coincidences between candidates in distinct astrophysical
channels could strongly support a multi-messenger discovery. 

Most LIGO analyses have required two detectors to identify candidate
gravitational wave events~\citep{babak2013searching}.  In Advanced LIGO's first
observing run, this requirement excluded nearly half of the available
data\footnote{Data that passes Category 1 data quality checks. These DQ cuts
eliminate $\sim 6\%$ of coincident time.} from
analysis~\citep{TheLIGOScientific:2016pea}. Previous compact binary coalescence
(CBC) searches using prototype LIGO~\citep{Allen:1999yt} and
TAMA300~\citep{Tagoshi:2000bz} data analyzed single detector time. In O1, the
PyCBC pipeline cataloged single detector triggers primarily for detector
characterization purposes, and the search for gravitational waves
associated with gamma-ray bursts~\citep{Abbott:2016cjt} also analyzed times
with one operating interferometer. In Advanced LIGO's second observing run,
GW170817 was first identified as a LIGO Hanford trigger by the GstLAL online
pipeline with an estimated false-alarm-rate of $\sim$ 1 / 9000
years~\citep{GCN}.  We include single interferometer data in our search, and we
assign significances to O1 single detector candidates for the first time,
although we note that others have previously suggested methods to rank these
candidates~\citep{cannon2015likelihood,Messick:2016aqy,Callister:2017urp}.

 1-OGC~\citep{nitz20181} recently provided a catalog of gravitational wave
candidates in O1 data obtained via the Gravitational Wave Open Science Center
(GWOSC)\footnote{\url{https://www.gw-openscience.org/}}~\citep{Vallisneri:2014vxa}. The
search presented here differs in several major ways. First, we target binary
neutron star systems exclusively by applying a mass model to increase
sensitivity to those systems~\citep{Cannon:2012zt,FongThesis}. Second, we use a
denser grid of template waveforms to minimize signal loss caused by the
discrete nature of the template bank~\citep{Owen:1995tm}. Third, we include an
additional \TotalSingleTime{} of single detector time in our analysis to
increase the analyzed time and improve the sensitivity of the search. Fourth,
we include additional coincident data that was not analyzed in 1-OGC. Finally,
we include all candidates with false-alarm-rates less than one per day in our
list and we provide BAYESTAR~\citep{Singer:2015ema} sky localization estimates
for each candidate to encourage multi-messenger followup surveys.

\section{Search Description}

We used the GstLAL-based inspiral pipeline to conduct a matched-filter
analysis~\citep{allen2012findchirp, cannon2012toward,
Messick:2016aqy,Sachdev:2019vvd,gstlal,lalsuite,gstreamer} of data provided by GWOSC and spanning
\ObsOneStart{} to \ObsOneEnd{}. GWOSC data is only available for times that
pass Category 1 data quality checks~\citep{TheLIGOScientific:2017lwt}, leaving
\CoincidentTime{} of coincident data and \TotalSingleTime{} of single detector
data. We exclude times known to have hardware injections from our analysis and
apply no additional data quality cuts.  Additional information on the data
quality and hardware injections is available via GWOSC. 

\subsection{Template bank}

Matched-filter based analyses correlate the data with a discrete bank of
template waveforms~\citep{owen1999matched, harry2009stochastic, Ajith:2012mn}
that model the gravitational wave emission of compact
binaries~\citep{Blanchet:1995ez,Buonanno:2009zt, ajith2007phenomenological}. The template bank used for this search was designed to maximize
sensitivity to BNS mergers with component masses and spins motivated by double
neutron star binary observations~\citep{Ozel:2012ax,thorsett1999neutron,
TheLIGOScientific:2017qsa}. For astrophysical reasons, we consider component
spins that are purely aligned or anti-aligned with the orbital angular
momentum, and we limit the dimensionless spin magnitude to be $\leq 0.05$~\citep{TheLIGOScientific:2016qqj}. We
model the component masses of our target population with a Gaussian
distribution where $\bar{m} = 1.33 M_\odot$, $\sigma = 0.05 M_\odot$~\citep{Ozel:2012ax}. We
consider three standard deviations in mass and transform coordinates from
component mass to chirp mass, $\mathcal{M} = (m_1 m_2)^{3/5} / (m_1 +
m_2)^{1/5}$, as the chirp mass is the primary parameter that affects the
gravitational wave signal~\citep{finn1993observing}. We broaden the resulting
chirp mass distribution to allow for statistical errors in our measurements and
we increase the mean of the distribution to account for redshift. This results
in a bank that covers detector frame chirp masses of $\mathcal{M} \in (1.04
M_\odot , 1.36 M_\odot)$. 
 
The final bank of \BankSize{} template waveforms was constructed with the
\texttt{TaylorF2} approximant and a minimum match of $.99$, which ensures that
signals with arbitrary parameters have a \BankMM{} match with at least one
template in the bank. This high precision limits the loss of signals due to
using a discrete template bank to $\sim 3 \%$; previous searches in O1 data have used template banks that allowed signal loss up to $\lesssim 10\%$~\citep{DalCanton:2017ala,Mukherjee:2018yra,nitz20181}.

\begin{figure} \includegraphics[width=\columnwidth]{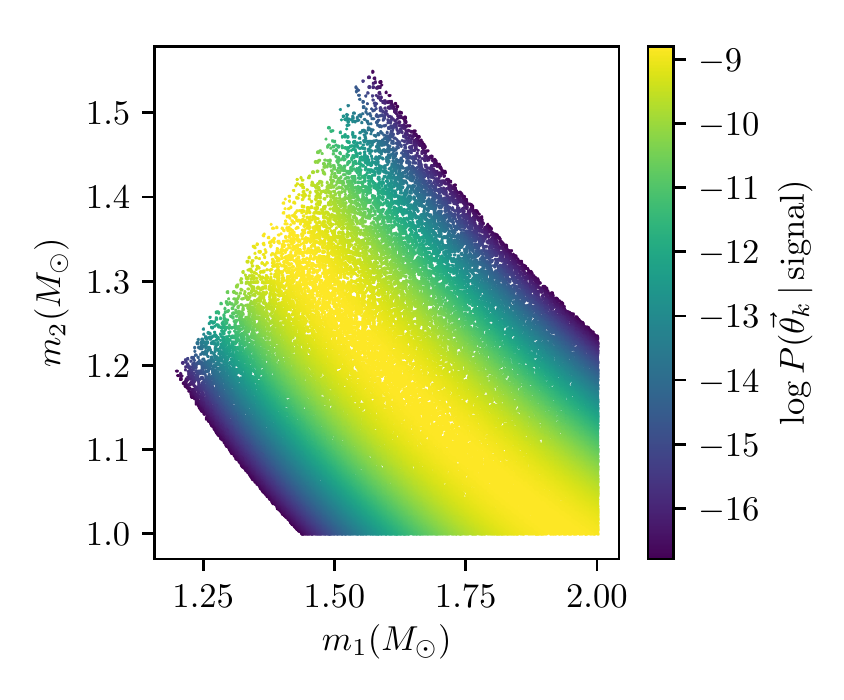}
\caption{\label{fig:bank} The template bank used for this search as depicted in
component masses, $m_1, m_2$, where $m_1 > m_2$. The colors represent the logarithm of probability
that a signal is recovered by a template $t_k$ (with parameters $\vec\theta$);
for this search, we have chosen a BNS population model with
a mean mass of $\bar{m} = 1.33M_\odot$ and a standard deviation of
$\sigma=0.05M_\odot$. The population model considers three
standard deviations in chirp mass. Although this population model
neglects effects due to redshift, redshift effects are considered when
we estimate the sensitivity of the search.}
\end{figure}

\subsection{Estimating significance of events}

We use a likelihood-ratio statistic, \likelihood{}, to rank candidate events by
their SNR, an autocorrelation based signal consistency check, the sensitivity
of each detector at the time of the candidate, and the time and phase delays
between different gravitational wave observatories~\citep{cannon2008bayesian,Cannon:2012zt,Dent:2013cva,
cannon2015likelihood,Messick:2016aqy,Hanna:2019ezx,Sachdev:2019vvd}. In addition we include an astrophysical
prior, which provides the probability that a signal from a BNS source
population is recovered by a particular template in the
bank~\citep{FongThesis}. The template bank and the prior probabilities associated with each template are shown in Fig.~\ref{fig:bank}.

Candidate events are assigned a false-alarm-rate that describes how often a
candidate with a likelihood-ratio statistic at least as high as its own is
expected to occur; the false-alarm-rate thus acts as a measure of how often the
noise can be expected to produce a candidate with similar properties. The first
gravitational wave detections had an extremely low false-alarm-rate (less than
1 / 100,000 years).  Here we are interested in digging considerably deeper into
the noise probing events with false-alarm-rates as high as 1 / day.

To estimate the false-alarm-rate for candidate events, we use triggers not
found in temporal coincidence between the interferometers when both LIGO
detectors were operating to estimate the background of noise-like
events~\citep{Cannon:2012zt, 2015arXiv150404632C, Messick:2016aqy}.  Single
detector events also have their background estimated from the set of
non-coincident triggers found when both LIGO detectors were operating, which
amounts to \CoincidentTime{} of data. We estimate our background from the
\CoincidentTime{} of coincident data. When a single detector candidate has a
higher likelihood ratio than any candidate in the background, we bound its
false-alarm-rate to 1 / \CoincidentTime{}.

\subsection{Estimating the sensitivity of the search}

The search sensitivity was estimated via Monte Carlo methods. We first
generated a set of BNS signals arising from systems with parameters consistent
with local populations --- we chose a Gaussian distribution for component
masses with $m_{\mathrm{mean}} = 1.33 M_\odot, \sigma_{m} = 0.05 M_\odot$ and
an isotropic distribution for spin. The injected population was modeled to a redshift of
$z = 0.2$, and probed a space-time volume of \ProbedVT{}. We rejected
\NumRejections{} simulated signals which had SNRs below $3$ to reduce the
number of compute cycles. The remaining \NumInjections{} fake signals were
injected into the data and subsequently searched for by the detection pipeline.
At a given false-alarm-rate threshold, we estimate the overall sensitivity of
the search via: 

\begin{align} \langle VT \rangle = \langle VT_{\mathrm{injected}} \rangle
\frac{N_{\mathrm{recovered}}}{N_{\mathrm{total \, sims}}} \end{align} 
where $N_{\mathrm{recovered}}$ varies with the false-alarm-rate threshold. This
search is approximately 30\% more sensitive at the 1 / 100 year
threshold than the previous BNS search presented at the end of Advanced LIGO's
first observing run~\citep{Abbott:2016ymx}. The inclusion of single detector
time in our analysis leads to a $\sim 33\%$ improvement in our
estimated $\langle VT \rangle$ at the 1 / day level. 

\section{Results}

We find no unambiguous gravitational wave events, but we identify
\NumCandidates{} candidates with false-alarm-rates less than one per day. We
provide the time, SNR, and false-alarm-rate of each candidate in
Table~\ref{table:results}, as well as the probability that the candidate is
astrophysical in origin ($p_{a}$). We compute $p_{a}$ using FGMC
methods~\citep{2015PhRvD..91b3005F,cannon2015likelihood}. When the $p_{a}$ assigned to single
detector candidates via FGMC exceeds the estimated single detector $p_{a}$
bound in~\citep{Callister:2017urp}, we substitute the lower bound. The associated
source parameters and sky localization estimates obtained via
BAYESTAR~\citep{Singer:2015ema} are provided on the LIGO Document Control
Center at \url{\WebsiteURL}. 

\begingroup
\squeezetable
\begin{table*}[t]
\setlength{\tabcolsep}{4pt}
\centering
\begin{tabular}[t]{lccccccccc}
\toprule

Date & FAR (yr$^{-1}$) & SNR & $p_{a}$\\
\midrule
2015-09-14T18:35:13.66$^{H}$ & 145.45 & 8.59 & $3.75 \times 10^{-3}$\\
2015-09-18T06:38:39.21 & 261.92 & 8.04 & $2.18 \times 10^{-3}$\\
2015-09-18T22:47:27.39$^{\dagger}$ & 193.46 & 8.52 & $2.92 \times 10^{-3}$\\
2015-09-19T00:05:01.08 & 326.71 & 7.63 & $1.78 \times 10^{-3}$\\
2015-09-21T10:10:02.92$^{H}$ & 7.52 & 8.10 & $6.95 \times 10^{-2}$\\
2015-09-22T11:26:08.35 & 312.67 & 8.61 & $1.86 \times 10^{-3}$\\
2015-09-23T13:47:35.79 & 165.39 & 8.56 & $3.38 \times 10^{-3}$\\
2015-09-24T00:53:02.68 & 19.68 & 8.45 & $2.29 \times 10^{-2}$\\
2015-09-24T05:57:35.24 & 107.32 & 8.71 & $4.88 \times 10^{-3}$\\
2015-09-25T01:24:33.74 & 56.59 & 9.15 & $8.73 \times 10^{-3}$\\
2015-09-25T21:15:15.92 & 38.39 & 8.58 & $1.25 \times 10^{-2}$\\
2015-09-26T23:51:25.50 & 56.05 & 8.39 & $8.81 \times 10^{-3}$\\
2015-09-27T14:28:55.77 & 243.80 & 8.60 & $2.32 \times 10^{-3}$\\
2015-09-29T01:46:01.42$^{\dagger}$ & 251.32 & 8.50 & $2.26 \times 10^{-3}$\\
2015-09-29T12:25:33.33 & 358.03 & 8.64 & $1.62 \times 10^{-3}$\\
2015-10-01T00:21:02.89 & 293.57 & 8.70 & $1.97 \times 10^{-3}$\\
2015-10-01T05:32:40.37 & 15.49 & 8.94 & $2.83 \times 10^{-2}$\\
2015-10-02T01:49:03.99 & 118.27 & 9.21 & $4.49 \times 10^{-3}$\\
2015-10-02T04:01:03.45 & 190.83 & 8.99 & $2.96 \times 10^{-3}$\\
2015-10-04T22:32:11.75$^{H}$ & 30.52 & 8.17 & $1.53 \times 10^{-2}$\\
2015-10-05T07:12:04.91 & 104.11 & 8.46 & $5.02 \times 10^{-3}$\\
2015-10-05T22:29:34.31 & 139.59 & 8.24 & $3.88 \times 10^{-3}$\\
2015-10-09T23:08:05.70 & 292.60 & 8.19 & $1.98 \times 10^{-3}$\\
2015-10-12T02:40:22.39 & 142.27 & 8.42 & $3.82 \times 10^{-3}$\\
2015-10-12T14:26:43.18 & 322.93 & 8.35 & $1.80 \times 10^{-3}$\\
2015-10-13T14:29:57.73$^{H}$ & 37.36 & 9.02 & $1.28 \times 10^{-2}$\\
2015-10-14T05:29:42.91$^{\dagger}$ & 149.36 & 8.75 & $3.68 \times 10^{-3}$\\
2015-10-18T19:03:46.85$^{H}$ & 7.52 & 8.05 & $0.181$\\
2015-10-19T17:37:05.25 & 124.01 & 8.78 & $4.30 \times 10^{-3}$\\
2015-10-24T09:01:50.34$^{L}$ & 94.09 & 10.56 & $5.53 \times 10^{-3}$\\
2015-10-24T09:03:52.00$^{L}$ & 7.52 & 9.69 & $7.96 \times 10^{-2}$\\
2015-10-24T19:53:05.66 & 360.26 & 8.57 & $1.61 \times 10^{-3}$\\
2015-10-28T12:24:31.67$^{H}$ & 7.52 & 9.06 & $0.181$\\
2015-10-28T17:03:45.19$^{\dagger}$ & 16.08 & 8.82 & $2.74 \times 10^{-2}$\\
2015-10-28T17:05:21.17$^{\dagger}$ & 0.78 & 10.63 & $0.289$\\
2015-10-29T08:27:29.92 & 345.02 & 9.04 & $1.68 \times 10^{-3}$\\
2015-10-29T11:48:01.64 & 58.64 & 8.78 & $8.45 \times 10^{-3}$\\
2015-10-29T12:05:48.00 & 363.99 & 8.24 & $1.59 \times 10^{-3}$\\
2015-10-29T19:18:33.06 & 193.47 & 8.26 & $2.92 \times 10^{-3}$\\
2015-10-30T00:08:56.47 & 358.38 & 8.55 & $1.62 \times 10^{-3}$\\
2015-10-30T04:08:58.11 & 240.56 & 8.44 & $2.35 \times 10^{-3}$\\
2015-10-31T10:27:43.77 & 320.37 & 8.05 & $1.81 \times 10^{-3}$\\
2015-10-31T11:30:36.72 & 329.59 & 8.35 & $1.76 \times 10^{-3}$\\
2015-10-31T22:01:00.91$^{L}$ & 331.06 & 7.97 & $1.76 \times 10^{-3}$\\
2015-11-01T11:13:23.94$^{L}$ & 12.17 & 8.65 & $3.50 \times 10^{-2}$\\
2015-11-04T13:37:23.67$^{\dagger}$ & 103.50 & 8.43 & $5.05 \times 10^{-3}$\\
2015-11-04T15:16:09.12$^{\dagger}$ & 69.89 & 9.12 & $7.23 \times 10^{-3}$\\
2015-11-05T06:20:44.61 & 312.42 & 8.56 & $1.86 \times 10^{-3}$\\
2015-11-06T07:44:18.43 & 95.56 & 8.42 & $5.45 \times 10^{-3}$\\
2015-11-06T10:07:13.79$^{\dagger}$ & 172.79 & 8.55 & $3.25 \times 10^{-3}$\\
2015-11-06T11:05:19.24 & 211.28 & 9.18 & $2.67 \times 10^{-3}$\\
2015-11-06T22:32:34.11 & 190.79 & 8.33 & $2.96 \times 10^{-3}$\\
\bottomrule
\end{tabular}
\begin{tabular}[t]{lccccccccc}
\toprule

Date & FAR (yr$^{-1}$) & SNR & $p_{a}$\\
\midrule
2015-11-10T00:32:55.28$^{\dagger}$ & 313.96 & 8.86 & $1.85 \times 10^{-3}$\\
2015-11-12T20:56:57.33 & 287.61 & 8.63 & $2.01 \times 10^{-3}$\\
2015-11-15T20:03:17.46 & 26.66 & 8.35 & $1.73 \times 10^{-2}$\\
2015-11-15T23:04:35.21 & 359.97 & 8.42 & $1.61 \times 10^{-3}$\\
2015-11-16T10:59:11.86 & 189.42 & 8.24 & $2.98 \times 10^{-3}$\\
2015-11-17T06:34:02.07$^{H}$ & 7.52 & 8.84 & $0.181$\\
2015-11-20T21:07:08.38$^{\dagger}$ & 15.60 & 8.95 & $2.81 \times 10^{-2}$\\
2015-11-21T22:26:44.55 & 104.06 & 8.65 & $5.02 \times 10^{-3}$\\
2015-11-26T13:34:13.65 & 6.09 & 8.68 & $6.23 \times 10^{-2}$\\
2015-11-28T08:29:19.80 & 229.16 & 8.16 & $2.46 \times 10^{-3}$\\
2015-11-28T14:05:27.32 & 128.85 & 8.55 & $4.16 \times 10^{-3}$\\
2015-11-29T03:39:34.71 & 250.42 & 9.27 & $2.27 \times 10^{-3}$\\
2015-12-02T10:45:49.81 & 201.50 & 9.24 & $2.80 \times 10^{-3}$\\
2015-12-02T15:17:48.11 & 308.63 & 9.28 & $1.88 \times 10^{-3}$\\
2015-12-02T17:38:00.95$^{\dagger}$ & 363.08 & 8.14 & $1.60 \times 10^{-3}$\\
2015-12-03T20:18:18.94 & 110.58 & 8.37 & $4.76 \times 10^{-3}$\\
2015-12-04T01:53:39.14 & 225.02 & 9.09 & $2.50 \times 10^{-3}$\\
2015-12-04T21:14:59.74$^{\dagger}$ & 8.89 & 9.04 & $4.57 \times 10^{-2}$\\
2015-12-05T10:16:47.45 & 284.26 & 8.59 & $2.03 \times 10^{-3}$\\
2015-12-06T06:50:38.17$^{L}$ & 77.45 & 7.72 & $6.64 \times 10^{-3}$\\
2015-12-08T09:27:47.71 & 344.81 & 8.27 & $1.68 \times 10^{-3}$\\
2015-12-08T13:22:36.24 & 47.36 & 8.76 & $1.03 \times 10^{-2}$\\
2015-12-09T07:25:24.68 & 141.65 & 7.85 & $3.84 \times 10^{-3}$\\
2015-12-14T18:15:44.85 & 145.53 & 8.43 & $3.75 \times 10^{-3}$\\
2015-12-14T19:32:20.42 & 145.58 & 8.72 & $3.75 \times 10^{-3}$\\
2015-12-15T06:04:29.41 & 20.34 & 8.49 & $2.23 \times 10^{-2}$\\
2015-12-15T10:53:01.22 & 154.61 & 8.78 & $3.58 \times 10^{-3}$\\
2015-12-18T00:56:19.12 & 83.80 & 8.19 & $6.19 \times 10^{-3}$\\
2015-12-18T09:59:11.16 & 147.23 & 8.71 & $3.72 \times 10^{-3}$\\
2015-12-20T05:33:58.81 & 300.99 & 7.86 & $1.92 \times 10^{-3}$\\
2015-12-22T10:08:48.42 & 234.05 & 9.22 & $2.41 \times 10^{-3}$\\
2015-12-23T00:07:10.93 & 18.95 & 8.99 & $2.36 \times 10^{-2}$\\
2015-12-23T12:23:35.72 & 60.11 & 10.25 & $8.26 \times 10^{-3}$\\
2015-12-23T13:50:49.48 & 178.46 & 8.00 & $3.16 \times 10^{-3}$\\
2015-12-23T16:13:55.82 & 290.02 & 8.98 & $1.99 \times 10^{-3}$\\
2015-12-24T23:05:56.58 & 47.49 & 10.08 & $1.03 \times 10^{-2}$\\
2015-12-24T23:06:28.51 & 146.99 & 9.55 & $3.72 \times 10^{-3}$\\
2015-12-24T23:06:57.04 & 70.65 & 9.42 & $7.16 \times 10^{-3}$\\
2015-12-25T02:16:31.87 & 320.05 & 8.49 & $1.82 \times 10^{-3}$\\
2015-12-28T21:04:05.90$^{H}$ & 160.93 & 8.57 & $3.46 \times 10^{-3}$\\
2015-12-29T11:50:15.09$^{H}$ & 234.41 & 8.23 & $2.41 \times 10^{-3}$\\
2015-12-31T11:20:54.32$^{H}$ & 180.00 & 8.82 & $3.13 \times 10^{-3}$\\
2016-01-02T02:47:29.35 & 356.13 & 7.51 & $1.63 \times 10^{-3}$\\
2016-01-02T02:54:39.60 & 239.65 & 8.11 & $2.36 \times 10^{-3}$\\
2016-01-03T02:29:54.78$^{\dagger}$ & 237.44 & 8.56 & $2.38 \times 10^{-3}$\\
2016-01-03T17:23:13.26 & 208.47 & 8.91 & $2.70 \times 10^{-3}$\\
2016-01-08T09:21:19.61 & 136.59 & 8.89 & $3.95 \times 10^{-3}$\\
2016-01-08T10:09:33.90$^{\dagger}$ & 218.62 & 8.52 & $2.58 \times 10^{-3}$\\
2016-01-12T05:19:01.34 & 107.14 & 8.34 & $4.89 \times 10^{-3}$\\
2016-01-15T08:37:05.94 & 328.35 & 8.19 & $1.77 \times 10^{-3}$\\
2016-01-19T05:40:13.04$^{\dagger}$ & 228.18 & 8.85 & $2.47 \times 10^{-3}$\\
\bottomrule
\end{tabular} \caption{\label{table:results}Binary neutron star triggers from
Advanced LIGO's first observing run with a false-alarm-rate (FAR) less than one
per day. We provide the time of coalescence, false-alarm-rate, SNR, and
astrophysical probability ($p_{a}$) for each candidate. Events marked by $H,L$
were found as single-detector triggers in LIGO-Hanford or LIGO-Livingston,
respectively. Events marked by a $\dagger$ occurred within 0.01 seconds of a
trigger in 1-OGC~\citep{nitz20181}. We expect $\mathcal{O}(1)$ of these
candidates to be gravitational waves. }
\end{table*}%
\endgroup

Although we cannot identify any one candidate from our list as astrophysical,
we can estimate the number of true signals buried in the list from our search sensitivity
and the expected binary neutron star merger rate. At a false-alarm-rate threshold of 1 / day, we
estimate $\langle VT \rangle = $ \VT{}. The LIGO Scientific Collaboration
recently estimated the local merger rate of binary neutron star systems to be
\BNSRate{} at 90\% confidence~\citep{LIGOScientific:2018mvr}; we adopt a
nominal value of $1000 \, \mathrm{Gpc}^{-3} \, \mathrm{yr}^{-1}$.  We therefore expect that
$ \langle VT \rangle \times \mathcal{R} =
0.67^{+2.0}_{-.60} $ of the candidates presented here are gravitational wave signals from binary neutron star coalescences. We stress that although the number of expected events depends on uncertainties
in both $\langle VT \rangle$ and $\mathcal{R}$, the expected number remains at
most $\mathcal{O}(1)$.

A single signal in our candidate list would imply a contamination fraction of 99\%. 
We provide the coalescence times in Table~\ref{table:results} and approximate
sky localizations online to encourage multi-messenger
searches that have the ability to illuminate true signals buried in the candidate list.

\section{Discussion}

We have presented a search for gravitational waves from BNS mergers.  Although
no gravitational wave signal was clearly identified in either single or double
interferometer time, we have provided a list of candidate events with
false-alarm-rates less than one per day.  The parameters for this search
overlap with those of gravitational wave catalogs
GWTC-1~\citep{LIGOScientific:2018mvr} and 1-OGC~\citep{nitz20181}. No shared
events are found between this candidate list and GWTC-1. While the GstLAL
pipeline identified a low-mass marginal candidate, 151012A, in GWTC-1, the
detector frame chirp mass is not covered by the bank used here. We note that
five single detector candidates meet the selection criteria
for inclusion in GWTC-1~\citep{LIGOScientific:2018mvr}.

For 1-OGC, we define overlapping candidates as those that share coalescence
times to a precision of two decimal places as differences between the pipelines
and the template banks can account for small differences in the measured time
of arrival. We find \OverlapOGC{} BNS candidates in common with 1-OGC. This is
not unexpected; 1-OGC has a trigger rate of $\sim 3000$ per day. They do not
assign any of the overlapping candidates a false-alarm-rate of less than one
per day. The variation in estimated false-alarm-rates can arise from
differences in the pipelines, template banks, and mass models used in the
searches.

In the hopes of enabling multi-messenger, sub-threshold follow-up, we have also
provided the coalescence times and sky localizations of the \NumCandidates{}
candidates with false-alarm-rates less than 1 / day. The analysis of single
detector time yielded \NumSingleCandidates{} of the \NumCandidates{} candidates
presented in our list, and nearly half of the analyzed data was obtained
during times at which only one detector was operating; this highlights the
importance of continuing to analyze single interferometer time in future
gravitational wave searches. 

\section{Acknowledgments}

We thank Peter Shawhan, Thomas Dent, and Jonah Kanner for useful feedback and discussion. This work was supported by the National Science Foundation through PHY-1454389,
OAC-1841480, PHY-1700765, and PHY-1607585. Computations for this research were
performed on the Pennsylvania State University’s Institute for CyberScience
Advanced CyberInfrastructure (ICS-ACI). This research has made use of data,
software and/or web tools obtained from the Gravitational Wave Open Science
Center (\url{https://www.gw-openscience.org}), a service of LIGO Laboratory,
the LIGO Scientific Collaboration and the Virgo Collaboration. LIGO is funded
by the U.S. National Science Foundation. Virgo is funded by the French Centre
National de Recherche Scientifique (CNRS), the Italian Istituto Nazionale della
Fisica Nucleare (INFN) and the Dutch Nikhef, with contributions by Polish and
Hungarian institutes. SRM thanks the LSSTC Data Science Fellowship Program,
which is funded by LSSTC, NSF Cybertraining Grant-1829740, the Brinson
Foundation, and the Moore Foundation. Funding for this project was provided by
the Charles E. Kaufman Foundation of The Pittsburgh Foundation. SC is supported
by the research programme of the Netherlands Organisation for Scientific
Research (NWO). HF acknowledges support from the Natural Sciences and
Engineering Research of Council of Canada (NSERC) and the Japan Society for the
Promotion of Science (JSPS). This paper has been assigned the document number
LIGO-P1800401.


\begin{thebibliography}{}
\expandafter\ifx\csname natexlab\endcsname\relax\def\natexlab#1{#1}\fi
\providecommand{\url}[1]{\href{#1}{#1}}

\bibitem[{Aasi {et~al.}(2015)}]{TheLIGOScientific:2014jea}
Aasi, J., {et~al.} 2015, Class. Quant. Grav., 32, 074001

\bibitem[{Abbott {et~al.}(2018{\natexlab{a}})}]{LIGOScientific:2018mvr}
Abbott, B., {et~al.} 2018{\natexlab{a}}, arXiv:1811.12907

\bibitem[{Abbott {et~al.}(2016{\natexlab{a}})}]{Abbott:2016blz}
Abbott, B.~P., {et~al.} 2016{\natexlab{a}}, Phys. Rev. Lett., 116, 061102

\bibitem[{Abbott {et~al.}(2016{\natexlab{b}})}]{Abbott:2016nmj}
---. 2016{\natexlab{b}}, Phys. Rev. Lett., 116, 241103

\bibitem[{Abbott {et~al.}(2016{\natexlab{c}})}]{TheLIGOScientific:2016pea}
---. 2016{\natexlab{c}}, Phys. Rev., X6, 041015

\bibitem[{Abbott {et~al.}(2016{\natexlab{d}})}]{Abbott:2016ymx}
---. 2016{\natexlab{d}}, Astrophys. J., 832, L21

\bibitem[{Abbott {et~al.}(2016{\natexlab{e}})}]{TheLIGOScientific:2016qqj}
---. 2016{\natexlab{e}}, Phys. Rev., D93, 122003

\bibitem[{Abbott {et~al.}(2017{\natexlab{a}})}]{GBM:2017lvd}
---. 2017{\natexlab{a}}, Astrophys. J., 848, L12

\bibitem[{Abbott {et~al.}(2017{\natexlab{b}})}]{Abbott:2016cjt}
---. 2017{\natexlab{b}}, Astrophys. J., 841, 89

\bibitem[{Abbott {et~al.}(2017{\natexlab{c}})}]{TheLIGOScientific:2017qsa}
---. 2017{\natexlab{c}}, Phys. Rev. Lett., 119, 161101

\bibitem[{Abbott {et~al.}(2018{\natexlab{b}})}]{TheLIGOScientific:2017lwt}
---. 2018{\natexlab{b}}, Class. Quant. Grav., 35, 065010

\bibitem[{Ajith {et~al.}(2014)Ajith, Fotopoulos, Privitera, Neunzert, \&
  Weinstein}]{Ajith:2012mn}
Ajith, P., Fotopoulos, N., Privitera, S., Neunzert, A., \& Weinstein, A.~J.
  2014, Phys. Rev., D89, 084041

\bibitem[{Ajith {et~al.}(2007)Ajith, Babak, Chen, Hewitson, Krishnan, Whelan,
  Bruegmann, Diener, Gonzalez, Hannam, {et~al.}}]{ajith2007phenomenological}
Ajith, P., Babak, S., Chen, Y., {et~al.} 2007, Classical and Quantum Gravity,
  24, S689

\bibitem[{Allen {et~al.}(2012)Allen, Anderson, Brady, Brown, \&
  Creighton}]{allen2012findchirp}
Allen, B., Anderson, W.~G., Brady, P.~R., Brown, D.~A., \& Creighton, J.~D.
  2012, Physical Review D, 85, 122006

\bibitem[{Allen {et~al.}(1999)}]{Allen:1999yt}
Allen, B., {et~al.} 1999, Phys. Rev. Lett., 83, 1498

\bibitem[{Babak {et~al.}(2013)Babak, Biswas, Brady, Brown, Cannon, Capano,
  Clayton, Cokelaer, Creighton, Dent, {et~al.}}]{babak2013searching}
Babak, S., Biswas, R., Brady, P., {et~al.} 2013, Physical Review D, 87, 024033

\bibitem[{Blanchet {et~al.}(1995)Blanchet, Damour, Iyer, Will, \&
  Wiseman}]{Blanchet:1995ez}
Blanchet, L., Damour, T., Iyer, B.~R., Will, C.~M., \& Wiseman, A. 1995, Phys.
  Rev. Lett., 74, 3515

\bibitem[{Buonanno {et~al.}(2009)Buonanno, Iyer, Ochsner, Pan, \&
  Sathyaprakash}]{Buonanno:2009zt}
Buonanno, A., Iyer, B., Ochsner, E., Pan, Y., \& Sathyaprakash, B.~S. 2009,
  Phys. Rev., D80, 084043

\bibitem[{Burns {et~al.}(2018)}]{Burns:2018pcl}
Burns, E., {et~al.} 2018, arXiv:1810.02764

\bibitem[{Callister {et~al.}(2017)Callister, Kanner, Massinger, Dhurandhar, \&
  Weinstein}]{Callister:2017urp}
Callister, T.~A., Kanner, J.~B., Massinger, T.~J., Dhurandhar, S., \&
  Weinstein, A.~J. 2017, Class. Quant. Grav., 34, 155007

\bibitem[{Cannon {et~al.}(2013)Cannon, Hanna, \& Keppel}]{Cannon:2012zt}
Cannon, K., Hanna, C., \& Keppel, D. 2013, Phys. Rev., D88, 024025

\bibitem[{Cannon {et~al.}(2015)Cannon, Hanna, \&
  Peoples}]{cannon2015likelihood}
Cannon, K., Hanna, C., \& Peoples, J. 2015, arXiv preprint arXiv:1504.04632

\bibitem[{{Cannon} {et~al.}(2015){Cannon}, {Hanna}, \&
  {Peoples}}]{2015arXiv150404632C}
{Cannon}, K., {Hanna}, C., \& {Peoples}, J. 2015, arXiv e-prints,
  arXiv:1504.04632

\bibitem[{Cannon {et~al.}(2012)Cannon, Cariou, Chapman, Crispin-Ortuzar,
  Fotopoulos, Frei, Hanna, Kara, Keppel, Liao, {et~al.}}]{cannon2012toward}
Cannon, K., Cariou, R., Chapman, A., {et~al.} 2012, The Astrophysical Journal,
  748, 136

\bibitem[{Cannon(2008)}]{cannon2008bayesian}
Cannon, K.~C. 2008, Classical and Quantum Gravity, 25, 105024

\bibitem[{Dal~Canton \& Harry(2017)}]{DalCanton:2017ala}
Dal~Canton, T., \& Harry, I.~W. 2017, arXiv:1705.01845

\bibitem[{Dent \& Veitch(2014)}]{Dent:2013cva}
Dent, T., \& Veitch, J. 2014, Phys. Rev., D89, 062002

\bibitem[{Essick {et~al.}(2017)}]{GCN}
Essick, R., {et~al.} 2017.
\newblock \url{https://gcn.gsfc.nasa.gov/other/G298048.gcn3}

\bibitem[{{Farr} {et~al.}(2015){Farr}, {Gair}, {Mandel}, \&
  {Cutler}}]{2015PhRvD..91b3005F}
{Farr}, W.~M., {Gair}, J.~R., {Mandel}, I., \& {Cutler}, C. 2015, \prd, 91,
  023005

\bibitem[{Finn \& Chernoff(1993)}]{finn1993observing}
Finn, L.~S., \& Chernoff, D.~F. 1993, Physical Review D, 47, 2198

\bibitem[{Fong(2018)}]{FongThesis}
Fong, H. K.~Y. 2018, PhD thesis, University of Toronto

\bibitem[{Gst{LAL}(2018)}]{gstlal}
Gst{LAL}. 2018, Gst{LAL} software: git.ligo.org/lscsoft/gstlal, ,

\bibitem[{Gstreamer(2018)}]{gstreamer}
Gstreamer. 2018, Gstreamer software: https://gstreamer.freedesktop.org, ,

\bibitem[{Hanna {et~al.}(2019)}]{Hanna:2019ezx}
Hanna, C., {et~al.} 2019, arXiv:1901.02227

\bibitem[{Harry {et~al.}(2009)Harry, Allen, \&
  Sathyaprakash}]{harry2009stochastic}
Harry, I.~W., Allen, B., \& Sathyaprakash, B. 2009, Physical Review D, 80,
  104014

\bibitem[{{LIGO Scientific Collaboration}(2018)}]{lalsuite}
{LIGO Scientific Collaboration}. 2018, {LIGO} {A}lgorithm {L}ibrary -
  {LALS}uite, free software (GPL), , , doi:10.7935/GT1W-FZ16

\bibitem[{Messick {et~al.}(2017)}]{Messick:2016aqy}
Messick, C., {et~al.} 2017, Phys. Rev., D95, 042001

\bibitem[{Mukherjee {et~al.}(2018)}]{Mukherjee:2018yra}
Mukherjee, D., {et~al.} 2018, arXiv:1812.05121

\bibitem[{Nitz {et~al.}(2018)Nitz, Capano, Nielsen, Reyes, White, Brown, \&
  Krishnan}]{nitz20181}
Nitz, A.~H., Capano, C., Nielsen, A.~B., {et~al.} 2018, arXiv preprint
  arXiv:1811.01921

\bibitem[{Owen(1996)}]{Owen:1995tm}
Owen, B.~J. 1996, Phys. Rev., D53, 6749

\bibitem[{Owen \& Sathyaprakash(1999)}]{owen1999matched}
Owen, B.~J., \& Sathyaprakash, B.~S. 1999, Physical Review D, 60, 022002

\bibitem[{Ozel {et~al.}(2012)Ozel, Psaltis, Narayan, \&
  Villarreal}]{Ozel:2012ax}
Ozel, F., Psaltis, D., Narayan, R., \& Villarreal, A.~S. 2012, Astrophys. J.,
  757, 55

\bibitem[{Sachdev {et~al.}(2019)}]{Sachdev:2019vvd}
Sachdev, S., {et~al.} 2019, arXiv:1901.08580

\bibitem[{Singer \& Price(2016)}]{Singer:2015ema}
Singer, L.~P., \& Price, L.~R. 2016, Phys. Rev., D93, 024013

\bibitem[{Smith {et~al.}(2013)Smith, Fox, Cowen, M{\'e}sz{\'a}ros,
  Te{\v{s}}i{\'c}, Fixelle, Bartos, Sommers, Ashtekar, Babu,
  {et~al.}}]{smith2013astrophysical}
Smith, M., Fox, D., Cowen, D., {et~al.} 2013, Astroparticle Physics, 45, 56

\bibitem[{Tagoshi {et~al.}(2001)}]{Tagoshi:2000bz}
Tagoshi, H., {et~al.} 2001, Phys. Rev., D63, 062001

\bibitem[{Thorsett \& Chakrabarty(1999)}]{thorsett1999neutron}
Thorsett, S.~E., \& Chakrabarty, D. 1999, The Astrophysical Journal, 512, 288

\bibitem[{Vallisneri {et~al.}(2015)Vallisneri, Kanner, Williams, Weinstein, \&
  Stephens}]{Vallisneri:2014vxa}
Vallisneri, M., Kanner, J., Williams, R., Weinstein, A., \& Stephens, B. 2015,
  J. Phys. Conf. Ser., 610, 012021

\end{thebibliography}

\end{document}